\documentstyle[prd,aps,floats]{revtex}
\begin{document}
\draft
%
%
\input epsf
\newcommand{\comm}[1]{}   \newcommand{\dcomm}[1]{#1}
\newcommand{\gsim}{\raisebox{-0.8ex}{\mbox{$\stackrel{\textstyle>}{\sim}$}}}
\newcommand{\lsim}{\raisebox{-0.8ex}{\mbox{$\stackrel{\textstyle<}{\sim}$}}}
\newcommand{\gl}{\raisebox{-0.8ex}{\mbox{$\stackrel{\textstyle >}{<}$}}}
\newcommand{\leg}{\raisebox{-0.8ex}{\mbox{$\stackrel{\textstyle <}{>}$}}}
\newcommand{\trh}{T_{\rm rh}}
\newcommand{\teff}{T_{\rm eff}}
\newcommand{\delcp}{\delta_{_{\rm CP}}}
\newcommand{\mh}{m_{_{\rm H}}}
\newcommand{\mw}{m_{_{\rm W}}}
\newcommand{\alphaw}{\alpha_{_{\rm W}}}
\newcommand{\ncs}{N_{_{\rm CS}}}

\renewcommand{\topfraction}{0.8}
\twocolumn[\hsize\textwidth\columnwidth\hsize\csname
@twocolumnfalse\endcsname
\preprint{IMPERIAL-TP-98/99-58, hep-ph/9912515}
\title{Inflaton-induced sphaleron transitions}
\author{Juan Garc{\'\i}a-Bellido}
\address{Theoretical Physics, Blackett Laboratory, Imperial College,
Prince Consort Road, London SW7 2BZ, U.K.}
\author{Dmitri Grigoriev}
\address{Institute for Nuclear Research of Russian Academy of
Sciences, Moscow 117312, Russia}
\date{December 23, 1999}
\maketitle

\begin{abstract}
It has recently been proposed that the production of long wavelength
Higgs and gauge configurations via parametric resonance at the end of
inflation may give rise to the required baryon asymmetry at the
electroweak scale.  We show that the stability of the inflaton
oscillations, long after the production of Higgs modes, keeps driving
the sphaleron transitions, which then become strongly correlated to the
inflaton oscillations. In models where the CP-violation operator is
related to time variations of the Higgs field, these correlations
immediately lead to an efficient generation of baryons that are not
washed out after the resonance.
\end{abstract}

\pacs{PACS number: 98.80.Cq, 11.30.Fs, 12.60.Fr, \ Preprint \ 
IMPERIAL-TP-99-58, hep-ph/9912515}

\vskip2pc]

\renewcommand{\thefootnote}{\arabic{footnote}}
\setcounter{footnote}{0}

\section{Introduction}

The origin of the matter-antimatter asymmetry in the universe is still
one of the most pressing problems of cosmology. Until recently it was
assumed that such an asymmetry could have arised at the electroweak
scale through a first order phase transition~\cite{krs,rs}, or via
leptogenesis~\cite{leptogen} at a much higher temperature. Recently, a
new mechanism for electroweak baryogenesis was proposed ~\cite{bgks},
based on the non-perturbative and out of equilibrium production of
long-wavelength Higgs and gauge configurations via parametric resonance
at the end of inflation. Such mechanism can be very efficient in
producing the required sphaleron transitions that gave rise to the
baryon asymmetry of the universe, in the presence of a CP-violating
interaction.

The new scenario~\cite{bgks} considers a very economical extension of
the symmetry breaking sector of the Standard Model with the only
inclusion of a singlet scalar field $\sigma$ that acts as an
inflaton.\footnote{This field is not necessarily directly related to the
inflaton field responsible for the temperature anisotropies of the
microwave background.} Its vacuum energy density drives a short period
of expansion, diluting all particle species, and its coupling to the
Higgs $\phi$ triggers the electroweak symmetry breaking. After
inflation, the inflaton oscillations induce resonant Higgs production,
via parametric resonance~\cite{KLS,GBL}, and out of equilibrium
sphaleron transitions.

One of the major problems that afflicted previous scenarios of
baryogenesis at the electroweak scale is the inevitability of a strong
wash-out of the generated baryons after the end of the CP-violation
stage during the phase transition. This problem was partially solved in
the scenario of Ref.~\cite{bgks}, where CP violation and efficient
topological (sphaleron) transitions coexist on roughly the same time
scale, during the resonant stage of preheating, while after-resonance
transitions are rapidly suppressed due to the decay of the Higgs and
gauge bosons into fermions and their subsequent thermalization below 100
GeV.

An important peculiarity of the new scenario is that it is possible for
the inflaton condensate to remain essentially spatially homogeneous for
many oscillation periods, even after the Higgs field has been produced
over a wide spectrum of modes. These inflaton oscillations induce a
coherent oscillation of the Higgs vacuum expectation value (VEV) through
its coupling to the inflaton, and thus induce possible CP-violating
interactions arising from operators containing the Higgs field. These
oscillations affect the sphaleron transition rate $\Gamma$ as well,
since the Higgs VEV determines the height of the sphaleron barrier,
therefore producing strong time correlations between variations in the
rate $\Gamma$ and the sign of CP violation.

We will show in this paper that such correlations can lead to
counterintuitive dynamical effects responsible for the steady
generation of baryons whenever sphaleron transitions occur, thus
making the usual post-resonance wash-out unlikely, independently of
the fermionic sector of the theory.

The key points are numerically illustrated with the 
(1+1)-dimensional Abelian Higgs toy model, extended with a neutral
(singlet) inflaton field, and a CP-violating operator~\cite{bgks}
\begin{eqnarray}
{\cal L} &=& - {1\over4}F_{\mu\nu}^2 - \kappa |\phi|^2 \,
\epsilon_{\mu\nu}F^{\mu\nu} \nonumber\\
 &+& |D_\mu\phi|^2 - {\lambda\over 4}(|\phi|^2 - v^2)^2 \label{toy} \\
 &+& {1\over 2} (\partial_\mu\sigma)^2 
- {1\over 2}g^2 \sigma^2 |\phi|^2\,. \nonumber
\end{eqnarray}
As described elsewhere~\cite{BS,GR,grsnpb,1p1}, this toy model contains
all the necessary ingredients to study electroweak baryogenesis. Note,
however, that we have not included the chiral fermions in the
discussion.\footnote{For a discussion of fermions in (1+1) models, see
Ref.~\cite{AS}.} We will assume for the moment that their coupling is
weak enough that only after many inflaton oscillation the decay of the
Higgs into fermions is relevant. The following discussion can be
extended directly to the (3+1)-dimensional case.
 
\begin{figure}[t]
\centering
\hspace*{-7mm}
\leavevmode\epsfysize=6.6cm \epsfbox{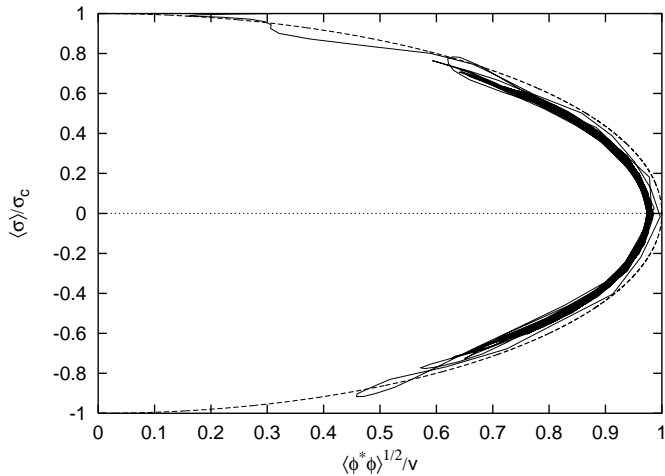}\\[3mm]
\caption{\label{fig2} The inflaton and Higgs zero modes coherently
oscillate near the trajectory $\sigma^2/\sigma_c^2+\phi^*\phi/v^2=1$
(dashed line).}
\end{figure}

It is also worth noticing that for a reasonably small $\kappa\,\lsim\,
0.1$, the CP-violating term in (\ref{toy}) does not interfere with the
dynamics of the inflaton and its ability to induce correlated sphaleron
transitions. This means that our results are applicable to any field
theory model with a $\phi^*\phi$-dependent CP-violation operator.

\section{Baryogenesis after hybrid inflation}

In the hybrid model of inflation considered in Ref.~\cite{bgks}, the
effective Higgs mass vanishes at the end of inflation, due to its
coupling to the inflaton field. This determines the inflaton amplitude
at this moment:
\begin{equation}\label{crit}
g^2\sigma^2_c = \lambda v^2 = M^2_H\,,
\end{equation}
where $M_H$ is the Higgs mass in the true vacuum, $\sigma~=~0,
\phi=v$. Due to the inflaton-Higgs coupling, the behaviour of these
fields during preheating after inflation is non standard; see
Ref.~\cite{GBL}. The inflaton field is dominated by its homogeneous zero
mode, which results in both Higgs and inflaton field coherently
oscillating close to the trajectory along the minimum of the potential,
$\sigma^2/\sigma_c^2+\phi^*\phi/v^2=1$; see Fig.~\ref{fig2}. This
behaviour holds for many inflaton oscillations, even after the higher
momentum modes of the Higgs field become populated via parametric
resonance and rescattering. As a consequence, the effective Higgs mass
and VEV are modified during that stage,

\begin{figure}[t]
\centering
\hspace*{-4.5mm}
\leavevmode\epsfysize=6.1cm \epsfbox{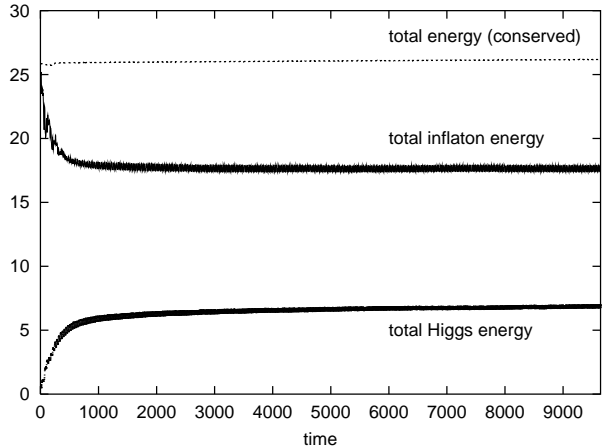}\\[3mm]
\caption{\label{fig1a} The time evolution of the inflaton and Higgs
  energies in the case of an incomplete resonance (in dimensionless
  units). The Higgs acquires here only $\sim 1/3$ of the initial energy,
  while the inflaton zero-momentum mode retains the remaining $2/3$. For
  a detailed discussion of field spectra, units of measure and parameter
  values see Ref. \protect\cite{gbg}.}
\end{figure}

\begin{eqnarray}
\tilde{v}^2 &=& v^2 - {g^2\over \lambda} \langle\sigma^2\rangle = 
v^2 (1 - \langle\sigma^2\rangle/\sigma_c^2) \label{v} \\
\tilde{M}^2_H &=& \lambda \tilde{v}^2 = M^2_H (1 -
\langle\sigma^2\rangle/\sigma_c^2) \label{M}
\end{eqnarray}
and the sphaleron mass, i.e. the height of the energy barrier, becomes
(in (1+1) dimensions, see Refs.~\cite{BS,GR})
\begin{equation}\label{msph}
E_{\rm sph} = {4 \over 3} \sqrt{\lambda} \tilde{v}^3 =
{4 \over 3} \sqrt{\lambda} v^3\left(1 -
{\langle\sigma^2\rangle\over\sigma_c^2}\right)^{3/2}\,,
\end{equation}
Note that Eq.(\ref{msph}) holds only at the maxima and minima of
$\langle\sigma^2\rangle$, when the Higgs potential is stationary.

\begin{figure}[tbp]
\centering
\hspace*{-4.5mm}
\leavevmode\epsfysize=6.1cm \epsfbox{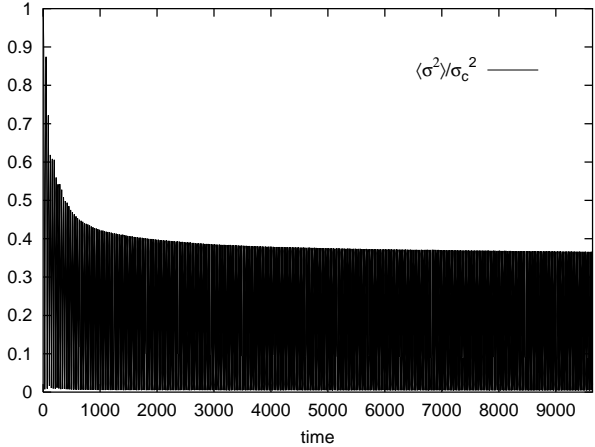}\\[3mm]
\caption{\label{fig1b} The time evolution of $\langle\sigma^2\rangle$.
After the resonance ends (at $t\sim 1000$), the inflaton field stays
almost completely homogeneous and keeps oscillating with an amplitude
comparable to the initial one, $\sigma_c=10$.}
\end{figure}

Large coherent oscillations of the inflaton field naturally occur in the
course of parametric resonance. Moreover, if the resonance doesn't
result in a complete decay of the inflaton (see Fig.~\ref{fig1a}), in
models without fermions, the inflaton keeps oscillating for a long
period of time (Fig.~\ref{fig1b}), limited only by bosonic
thermalization processes (which can be very
slow~\cite{heinz,PR,KT,gq98}) and the expansion of the Universe. As a
consequence, one expects substantial periodic variations in the height
of the energy barrier~(\ref{msph}) separating different topological
vacua. For stationary fields the barrier height equals the sphaleron
mass (\ref{msph}) and its variations are close to a factor of
$2^{3/2}$. In the low-temperature broken phase the topological
transition rate exponentially depends on the barrier height, so these
variations may result in large variations of the rate itself, especially
if the exponential suppression factor is large.

Once $\langle\phi^*\phi\rangle$ keeps oscillating below $v^2$, even
long after the beginning of the resonance, the periodical change in
the barrier height, see Eq.~(\ref{msph}), will result in a
corresponding increase of the sphaleron rate $\Gamma$, especially
noticeable in the latter stages, when $\Gamma$ becomes small, see
Fig.~14 of Ref.~\cite{bgks}. Initially the transitions are not
suppressed and hardly ever related to the barrier height. Note that
the energy transfer from the inflaton to the Higgs in the course of
parametric resonance will also result in a temporary increase in the
rate $\Gamma$, mostly at the beginning of the resonance,
see Ref.~\cite{bgks}.

\section{Inflaton induced sphaleron transitions}

An important consequence of the periodical variations in the barrier
height is that the probability of topological transitions will change
periodically in correlation with $\langle\phi^*\phi\rangle$ (see
Fig.~\ref{fig4}) and thus $\langle\sigma^2\rangle$ oscillations.  This
effect leads to interesting physical consequences for baryogenesis at
preheating~\cite{bgks}.

\begin{figure}[tbp]
\centering
\hspace*{-6mm}
\leavevmode\epsfysize=6.5cm \epsfbox{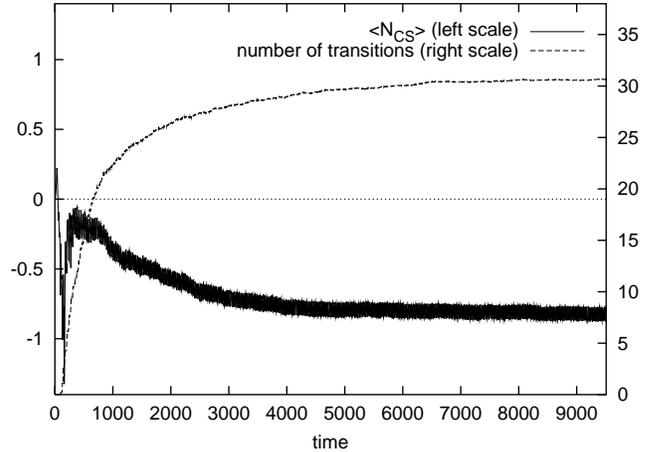}\\[3mm]
\caption{\label{fig3} The continuous production of baryons as a result
  of correlations between the topological transition rate and the
  CP violation generated by the term $ - \kappa\, \phi^*\phi\,
  \epsilon_{\mu\nu} F^{\mu\nu}$ in the Lagrangian. This plot corresponds
  to $\kappa=0.1$; for a detailed description see
  Ref. \protect\cite{bgks,gbg}. The solid line represents the shift in
  the Chern-Simons number, $\ncs$, averaged over an ensemble of a
  few hundred independent runs. The dashed line is the integral $\int
  \Gamma dt$, i.e. the average number of topological transitions
  accumulated per individual run.  Note the remarkable similarity of
  both curves for $t > 1000$. This means that all transitions at this
  stage are equally efficient in generating baryons, changing the
  Chern-Simons number by about $-1/20$ per transition for many
  oscillations, demonstrating the absence of baryon wash-out in the
  model.}
\end{figure}

\begin{figure}[tbp]
\centering
\hspace*{-7mm}
\leavevmode\epsfysize=6.6cm \epsfbox{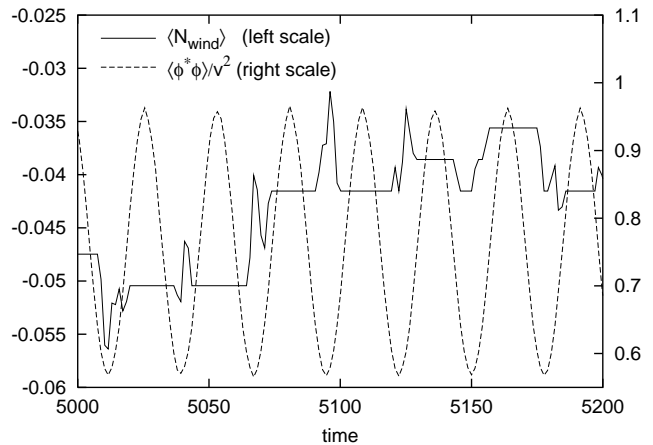}\\[3mm]
\caption{\label{fig4} The periodic variations of the Higgs effective
  potential also affect the sphaleron transitions, detected by changes
  in the Higgs winding number, $N_{\rm wind}$. This figure shows the
  average over an ensemble of a few hundred runs (solid line). In our
  model no transitions occur when $\langle\phi^*\phi\rangle$
  (dashed line) is close to its maximal value, which corresponds to
  the maximum height of the sphaleron barrier, $E_{\rm sph}$, see
  Eq.~(\ref{msph}), while the number of transitions is maximal when
  $\langle\phi^*\phi\rangle$ decreases to its minimum value. Of
  course, in each individual run $N_{\rm wind}$ has integer values.}
\end{figure}

In most field theoretical models with CP violation due to a
non-vanishing $\partial_0\langle\phi^*\phi\rangle$, such as the
two-Higgs model~\cite{1p1} discussed in the context of baryogenesis at
a thermal phase transition, the CP asymmetry is present, and baryons
are generated, only for a short period of time, as the Higgs field
moves to the new VEV through the phase transition. Unfortunately, in 
those models the generated asymmetry could be rapidly washed out by
late topological transitions.

\begin{figure}[t]
\centering
\hspace*{-6mm}
\leavevmode\epsfysize=6.5cm \epsfbox{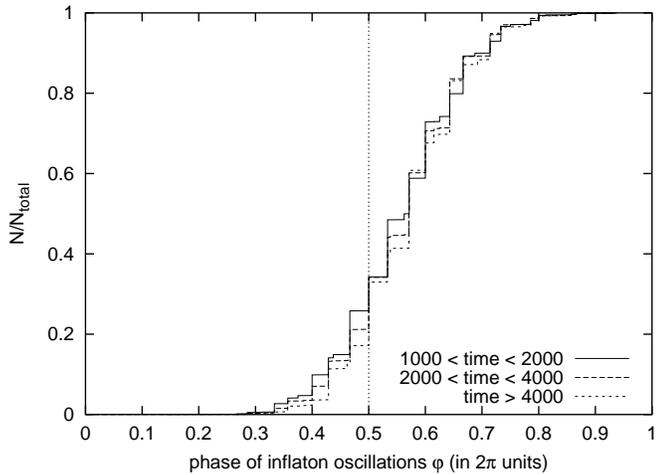}\\[3mm]
\caption{\label{fig5a} The distribution of sphaleron transitions as a
  function of the phase of inflaton oscillations (an integral histogram
  showing the ratio of transitions that happen at phases smaller than
  a given one). This figure shows that 75\% of all observed
  transitions took place near the maximum of $\langle\sigma^2\rangle$,
  i.e. when the inflaton phase is in the range $0.4<\varphi/2\pi<0.7$,
  and therefore at the minimum of $\langle\phi^*\phi\rangle$. Note
  that the distribution is slightly asymmetric -- more transitions
  happen after $\langle\phi^*\phi\rangle$ comes through its minimum
  (dotted line) -- probably due to the delayed relaxation of the
  winding number to its new stable value. However, the exact origin of
  this slight time delay is yet to be understood.}
\end{figure}

\begin{figure}
\centering
\hspace*{-6mm}
\leavevmode\epsfysize=6.5cm \epsfbox{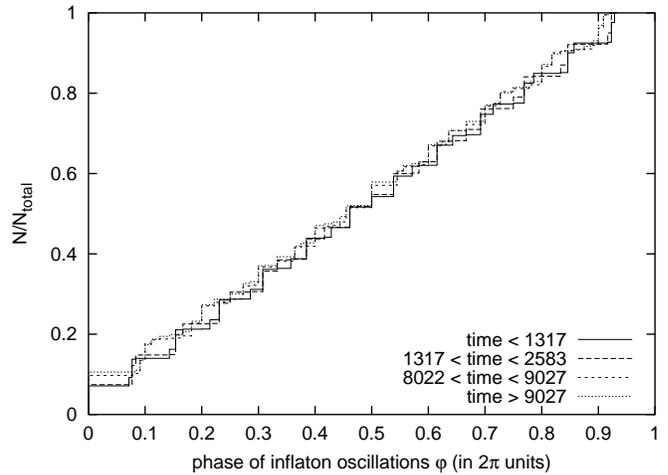}\\[3mm]
\caption{\label{fig5b} Analogous to Fig.~\ref{fig5a} but for the case 
  of a complete resonance; see Fig.4 of Ref.~\protect\cite{bgks}. This
  figure demonstrates the lack of correlations between $\Gamma$ and
  $\langle\sigma^2\rangle$ oscillations in this case. The observed
  transitions are equally distributed with the phase $\varphi$. }
\end{figure}

On the other hand, in the recent proposal~\cite{bgks} for electroweak
baryogenesis during preheating, the coherent long-term periodic
oscillations of the inflaton naturally lead to a long-lived CP
violation that shuts off only when the amplitude of oscillations
finally decreases with thermalization. The oscillations in
$\langle\phi^*\phi\rangle$ induce, via the CP-violating term in
Eq.~(\ref{toy}), an alternating chemical potential $\mu_{_{\rm eff}}
\propto -\kappa\,\partial_0\langle\phi^*\phi\rangle$. This means that
baryon and antibaryon production is biased in an alternating way with
each oscillation. Depending on the phase of $\langle\phi^*\phi\rangle$
oscillations, CP violation changes sign and vanishes after time
averaging. However, the topological transition rate is also changing
in accordance with precisely the same phase, as described above, thus
favouring CP violation of a certain sign -- see Fig.~\ref{fig3} --
that depends on the parameters and dynamics of the model; see
Ref.~\cite{gbg} for details. 

It is this correlation between CP violation and the growth in the rate
of sphaleron transitions which ensures that the baryonic asymmetry
generated is completely safe from wash-out, because of the long-term
nature of CP oscillations. Depending on initial conditions, the rate
$\Gamma$ can finally vanish, e.g. due to the (bosonic) thermalization
of the Higgs field, as seen in Fig.~\ref{fig3}, but this doesn't
affect the continuous pattern of CP-$\Gamma$ correlations. In other
words, these correlations effectively give rise to a permanent and
constant CP violation, thus preventing the generated asymmetry from
being washed out.

\begin{figure}[t]
\centering \hspace*{-8mm}
\leavevmode\epsfysize=6.5cm \epsfbox{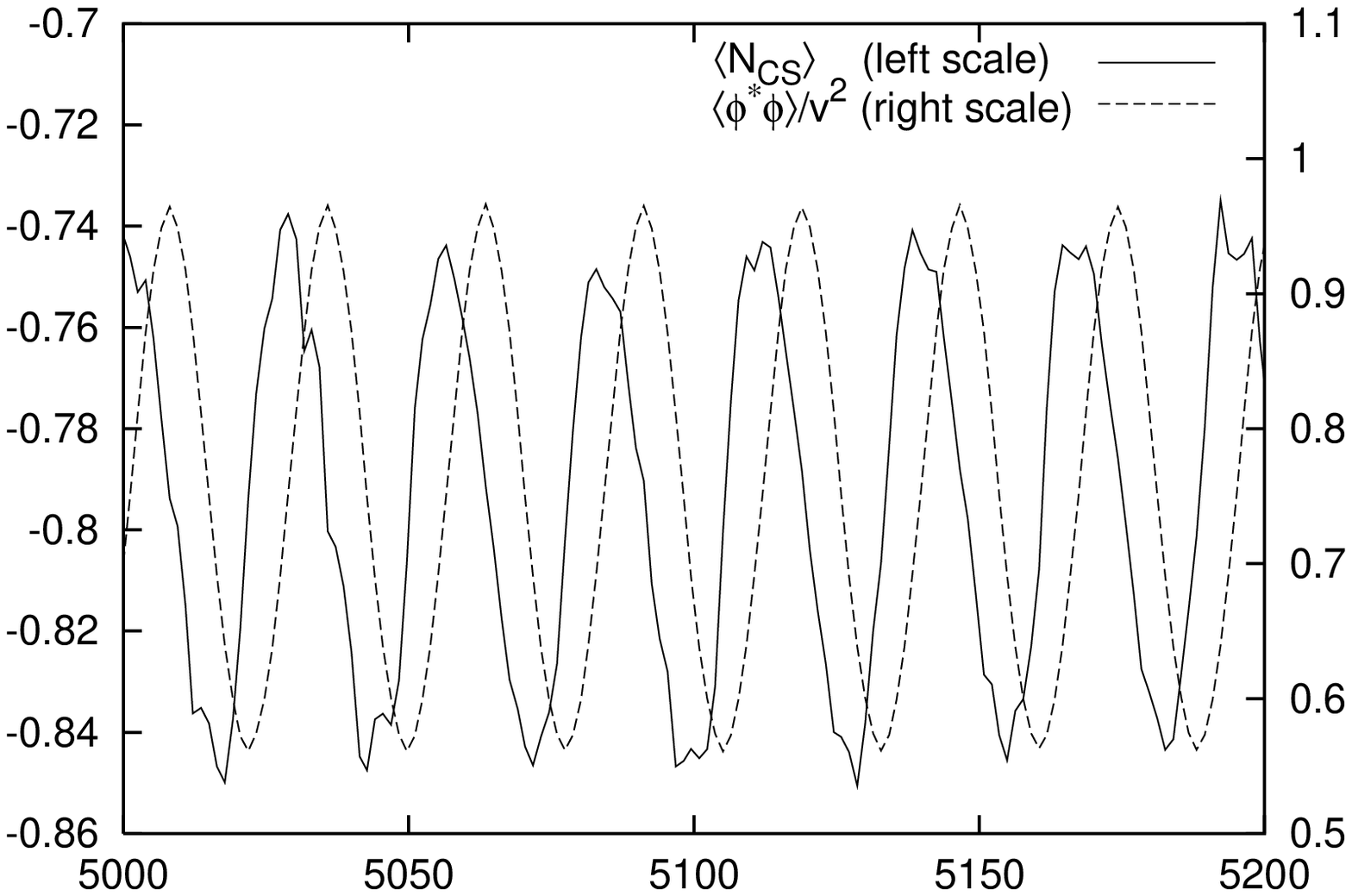}\\[3mm]
\hspace*{-8mm}
\leavevmode\epsfysize=6.5cm \epsfbox{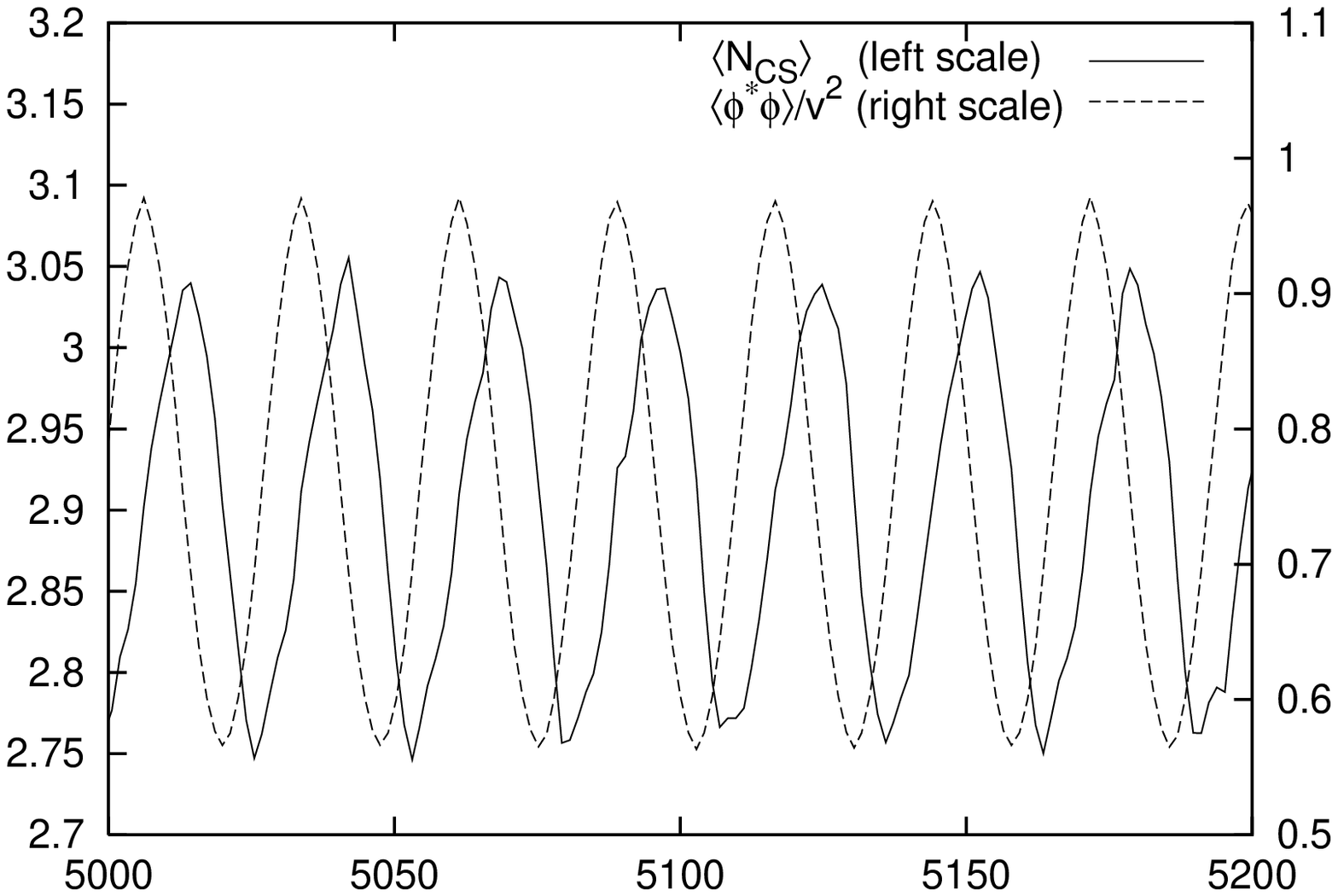}\\[3mm]
\caption{\label{fig6} In these plots we show the relation between the
periodical oscillations of Chern-Simons number $\ncs$ driven by the
CP-violating term and the Higgs oscillations that directly affect the
sphaleron transitions (as shown on Fig.~\ref{fig4}). The phase shift
between $\langle\ncs\rangle$ and $\langle\phi^*\phi\rangle$ determines
the sign of the permanent effective chemical potential which gives rise
to a continuous production of antibaryons ($\kappa=0.1$, higher plot and
Fig.~\ref{fig3}), and baryons ($\kappa=-0.25$, lower plot). Note that
this phase difference exactly equals to ${\pi\over 2}{\rm sign}\,\kappa$
and is insensitive to the absolute value of the CP-violation parameter.}
\end{figure}

The non-equilibrium and non-perturbative nature of these
inflaton-induced topological transitions makes their rigorous analysis
difficult. However, it is possible to study these transitions
numerically as in Ref.~\cite{bgks}, directly observing the time
correlations between the transition rate $\Gamma$ and the phase of
$\langle\phi^*\phi\rangle$ oscillations by comparing the corresponding
evolution plots, see Fig.~\ref{fig4}. A more detailed analysis can be
provided by the histogram distribution of the observed topological
transitions with the phase of $\langle\sigma^2\rangle$ oscillations,
Fig.~\ref{fig5a}. In this plot the phase $\varphi=0$ or $2\pi$
corresponds to $\langle\sigma^2\rangle$ being at its minimum and
$\langle\phi^*\phi\rangle$ at its maximum, and the phase $\varphi=\pi$
when the opposite is true. This figure clearly demonstrates that most
transitions occur when $\langle\phi^*\phi\rangle$ is close to its
minimum, where the sphaleron barrier is smaller. It also demonstrates
that the $\Gamma-\langle\sigma^2\rangle$ correlation remains valid
throughout the whole post-resonance period when (bosonic) thermalization
of the Higgs results in a gradual decrease of the rate $\Gamma$.

On the other hand, in the case of complete resonance, when
$\langle\sigma^2\rangle$ is small and the variations of
$\langle\phi^*\phi\rangle$ have no significant effect on the rate,
topological transitions become completely uncorrelated with the
inflaton oscillations, see Fig.~\ref{fig5b}.

\section{Conclusions}

It was shown in this paper that the presence of an oscillating inflaton
field coupled to the Higgs can considerably modify the dynamics of both
thermal and non-thermal topological transitions. Strong correlations
between a $\phi^*\phi$-dependent CP-violating operator and the rate of
sphaleron transitions $\Gamma$ generate a permanent (long-lived)
effective chemical potential driving baryogenesis; see
Fig.~\ref{fig6}. This extends the production of baryons for the whole
period during which sphaleron transitions occur, and disappears when the
latter vanishes, therefore preventing the wash-out stage.

\section*{Acknowledgements}

J.G.B. is supported by the Royal Society of London. D.G. is grateful to
Imperial College theory group for their kind hospitality. D.G. work at
IC was supported by the Royal Society. D.G. is also supported in part by
RBRF Grant No. 98-02-17493a.


\begin{thebibliography}{99}

\bibitem{krs} V.A.~Kuzmin, V.A.~Rubakov, and M.E.~Shaposhnikov,
Phys. Lett. {\bf 155B}, 36 (1985).

\bibitem{rs} For a review, see V.A.~Rubakov and M.E.~Shaposhnikov,
Phys. Usp. {\bf 39}, 461 (1996), {\tt hep-ph/9603208}.

\bibitem{leptogen} M.~Fukugita and T.~Yanagida, Phys. Lett. 
{\bf 174B}, 45 (1986).

\bibitem{bgks}  J.~Garc{\'\i}a-Bellido, D.~Grigoriev, A.~Kusenko and
M.~Shaposhnikov, Phys. Rev. D {\bf 60}, 123504 (1999), {\tt
hep-ph/9902449}.

\bibitem{KLS} L.~Kofman, A.~Linde and A.~A.~Starobinsky, Phys. Rev.
Lett. {\bf 73}, 3195 (1994), {\tt hep-th/9405187}; 
Phys. Rev. D {\bf 56}, 3258 (1997), {\tt hep-ph/9704452}.

\bibitem{GBL} J.~Garc\'\i a-Bellido and A.D.~Linde, Phys. Rev. D
{\bf 57}, 6075 (1998), {\tt hep-ph/9711360}.

\bibitem{BS} A.I.~Bochkarev and M.E.~Shaposhnikov,
Mod. Phys. Lett. {\bf A2}, 417 (1987).

\bibitem{GR} D.Yu.~Grigoriev and V.A.~Rubakov, Nucl. Phys. {\bf B299},
67 (1988).

\bibitem{grsnpb} D.Yu.~Grigoriev, V.A.~Rubakov and M.E.~Shaposhnikov,
  Phys. Lett. {\bf B216}, 172 (1989);
  Nucl. Phys. {\bf B326}, 737 (1989).

\bibitem{1p1}  D.Yu.~Grigoriev, M.E.~Shaposhnikov and N.G.~Turok,
Phys. Lett. {\bf B275}, 395 (1992).

\bibitem{AS} G. Aarts and J. Smit, Nucl. Phys. {\bf B555}, 355 (1999),
{\tt hep-ph/9812413}; Phys. Rev. D {\bf 61}, 025002 (2000), 
{\tt hep-ph/9906538}.

\bibitem{gbg}  J.~Garc{\'\i}a-Bellido and D.~Grigoriev, in
preparation.

\bibitem{heinz} U.~Heinz, C.R.~Hu, S.~Leupold, S.G.~Matinian and
B.~M\"uller, Phys. Rev. D {\bf 55}, 2464 (1997), {\tt hep-th/9608181}.

\bibitem{PR} T.~Prokopec and T.G.~Roos, Phys. Rev. D {\bf 55},
3768 (1997), {\tt hep-ph/9610400}.

\bibitem{KT} S.~Yu.~Khlebnikov and I.~I.~Tkachev, Phys. Rev. Lett.
{\bf 77}, 219 (1996), {\tt hep-ph/9603378}; 
Phys. Rev. Lett. {\bf 79}, 1607 (1997), {\tt hep-ph/9610477}.

\bibitem{gq98} D.Yu.~Grigoriev, in: Procs. of the 10th Int.  Seminar
QUARKS-98 ({\tt http://www.inr.ac.ru/\raisebox{-0.5ex}{\~}q98/proc/}).

\end{thebibliography}
\end{document}